\documentclass[11pt,a4paper]{article}
\pdfoutput=1

\usepackage[T1]{fontenc} 

\usepackage{color,xcolor,graphicx,amsmath,amssymb,mathtools,amsfonts,hyperref,cleveref,subcaption,booktabs,multirow,geometry,ulem,cancel,tablefootnote,physics}
\usepackage[export]{adjustbox}

\newcommand{\be}{\begin{eqnarray}}
\newcommand{\ee}{\end{eqnarray}}
\newcommand{\baln}{\begin{aligned}}
\newcommand{\ealn}{\end{aligned}}
\newcommand{\lt}{\left }
\newcommand{\rt}{\right }

\usepackage{auncial}
\usepackage[B1]{fontenc}

\definecolor{amber(sae/ece)}{rgb}{1.0, 0.49, 0.0}

\definecolor{blue(ncs)}{rgb}{0.0, 0.53, 0.74}

\usepackage{slashed}
\hypersetup{colorlinks,linkcolor={green!50!black},citecolor={magenta},urlcolor={magenta}}

\def\g{\gamma}

\definecolor{darkpastelgreen}{rgb}{0.01, 0.75, 0.24}
\usepackage{scalerel}
\usepackage{tikz}
\usetikzlibrary{svg.path}

\definecolor{orcidlogocol}{HTML}{A6CE39}
\tikzset{
  orcidlogo/.pic={
    \fill[orcidlogocol] svg{M256,128c0,70.7-57.3,128-128,128C57.3,256,0,198.7,0,128C0,57.3,57.3,0,128,0C198.7,0,256,57.3,256,128z};
    \fill[white] svg{M86.3,186.2H70.9V79.1h15.4v48.4V186.2z}
                 svg{M108.9,79.1h41.6c39.6,0,57,28.3,57,53.6c0,27.5-21.5,53.6-56.8,53.6h-41.8V79.1z M124.3,172.4h24.5c34.9,0,42.9-26.5,42.9-39.7c0-21.5-13.7-39.7-43.7-39.7h-23.7V172.4z}
                 svg{M88.7,56.8c0,5.5-4.5,10.1-10.1,10.1c-5.6,0-10.1-4.6-10.1-10.1c0-5.6,4.5-10.1,10.1-10.1C84.2,46.7,88.7,51.3,88.7,56.8z};
  }
}

\newcommand\orcidicon[1]{\href{https://orcid.org/#1}{\mbox{\scalerel*{
\begin{tikzpicture}[yscale=-1,transform shape]
\pic{orcidlogo};
\end{tikzpicture}
}{|}}}}

\usepackage{hyperref} 

\providecommand{\keywords}[1]{\small \textbf{\textit{Keywords---}} #1}

\usepackage{changepage}

\title{\bf {\textcolor{red}{
Particle dark matter density and entropy production in the early universe}}
}

\author{Arnab~Chaudhuri\thanks{corresponding author}\ 
$^{a}$\footnote{{\bf e-mail}: \href{mailto:arnabchaudhuri.7@gmail.com}{arnabchaudhuri.7@gmail.com}}%
\ \orcidicon{0000-0002-6784-1360},
Maxim Yu. Khlopov$^{b}$\footnote{{\bf e-mail:} \href{mailto:khlopov@apc.in2p3.fr}{khlopov@apc.in2p3.fr}}\ \orcidicon{0000-0002-1653-6964} and
Shiladitya Porey$^{a}$
\ \orcidicon{0000-0002-2460-6723}
\\
$^a$ \small{ Novosibirsk State University} \\
\small{ Pirogova ul., 2, 630090 Novosibirsk, Russia}\\
$^b$ 
\small{CNRS, Astroparticule et Cosmologie, Université de Paris, F-75013 Paris, France; 
}\\
\small{Institute of Physics, Southern Federal University, 344090 Rostov on Don, Russia}\\
\small{Center for Cosmopartilce Physics “Cosmion”}\\
\small{National Research Nuclear University “MEPhI” (Moscow~Engineering Physics Institute)}\\
\small{115409 Moscow, Russia}
}

\begin{document}

\maketitle

\begin{abstract}
Dark Matter (DM) density is reduced if entropy production takes place after {DM} particles abundance is frozen out in the early universe. We study a possibility of such reduction due to entropy production in the electroweak phase transition (EWPT). We compare scenarios of entropy production in the standard model (SM) and its simplest extension, the two-Higgs doublet model (2HDM). Assuming the EWPT is of second order in the SM scenario and the first order in the 2HDM, we calculate the entropy release in these scenarios and the corresponding dilution of preexisting {DM} density in the early universe. We find the effect of dilution in EWPT  significant for confrontation with observations of any form of possible {DM} (including primordial black holes (PBHs)), which is frozen out, decoupled, frozen in, or formed before EWPT.
\end{abstract}


\section{Introduction}

The early universe is associated with the high temperature and density of particles. In accordance with the Big Bang theory, the temperature during the very early stages of expansion ranged from few {$eV$} 
 to Planck scale $\sim M_{Pl}$~$\sim 10^{19}$~{$GeV$} 
 (more precisely to reheating temperature in inflationary models). The physical basis of now standard cosmological paradigm of the inflationary universe with baryosynthesis and { Dark Matter (DM)}/energy involves extensions of the particle Standard Model (SM) and depending on the symmetry breaking pattern of the Beyond the Standard model (BSM) physics may include several phase transitions in the cosmological scenario. However, any scenario based on BSM physics should inevitably include phase transitions predicted by SM, and first of all, related to breaking of the electroweak symmetry. 

The period of the universe before breaking of electroweak symmetry, when electromagnetic and weak force  remained merged into a single electroweak force is generally known as the electroweak epoch
~\cite{1,2}. The electroweak epoch ended with electroweak phase transition (EWPT). Depending on the parameters of the BSM model, the nature of the phase transition can be of the first order, second order, or it can be a smooth crossover. The first order EWPT can be a source of gravitational waves~\cite{3,4}. EWPT, in general, cannot only be a potential candidate for baryogenesis (if the three Sakharov's conditions~\cite{5} are satisfied) but also it can be a source to dilute the preexisting baryon asymmetry as well as of any form of {DM}, which is frozen out, decoupled, frozen in or formed
 before the EWPT.

With the proviso that thermal equilibrium is sustained, the entropy of the universe should remain conserved according to the second law of thermodynamics. In this article we are engrossed in the earliest epochs of the universe such that
Bosonic and fermionic species were massless and  with negligible chemical potential
, as mentioned in the textbooks ~\cite{Gorbunov:2011zzc, Bambi:2015mba}, 
relative to the temperature of the universe $T(t)$. The above assumption is accurate just before the EWPT and this with the condition of conservation of energy density of plasma lead to entropy conservation law: 
\be \label{ent-consv}
s=\frac{\mathcal{P}+\rho}{T}a^3=const.
\ee
where $s$ is the entropy density per comoving volume, $a(t)$ is the scale factor, $T(t)$ is the temperature of the fluid (or plasma), $\rho$ and $P$ are the plasma energy density and pressure, respectively.

To attain the state of local thermal equilibrium and the maximum entropy state, the interaction rate $\Gamma$ among the ultra-relativistic particles in the hot dense plasma must be greater than Hubble expansion rate, ${\cal H}$. $\Gamma =\sigma n v$, the product of cross section among particles, their number density and relative velocity ($v\sim 1$ for photons) and thus it can be stated that for ${\cal O}(1)$ estimation, $\Gamma \simeq \alpha^n T(t)$ ($n=1$ for decay and $2$ for two-body scattering). Under radiation domination, the ${\cal H}$ varies as ${\cal H}\sim T(t)^2/ M_{Pl}$, $M_{Pl}$ being the reduced Planck mass. The condition $\Gamma> {\cal H}$ implies $T< \alpha^n M_{Pl}$. This relation was verifiable in the very early universe because $\alpha \sim 10^{-2}$ and varies logarithmically with $T(t)$~\cite{Chaudhuri:2021agl}.


In thermal equilibrium, the distribution function of any species is defined by their chemical potential ($\mu_j$) and the temperature ($T$) of the plasma. The temperature of the plasma is similar to all the species. The equilibrium distribution function of the species are represented by the Bose-Einstein distribution and Fermi-Dirac distribution for bosons
and fermions, respectively, with $\mu_j=0$ and has the following form:
\be
f(E)=\frac{1}{exp(E/T \pm 1)}.
\ee

Entropy density is of our special interest because it helps us to estimate the number density of any particle per comoving volume and more entropy influx implies dilution of the number density of any particle. The number density of the frozen out species are conserved in the early universe but there can be some periods when their dilution is possible due to the influx of entropy into the plasma. If the energy density of the universe was dominated by that of the primordial black holes of sufficient masses such that they evaporate in the corresponding early matter-dominated epoch, the influx of entropy can sufficiently reduce the pre-existing baryon asymmetry and {DM} density, if they were created before this epoch (see~\cite{Dolgov:2000ht, Chaudhuri:2020wjo}). {Quantum chromodynamics (QCD)} phase transition at \mbox{$T \sim$ 150--200~{$GeV$}} can also lead to such dilution as well~\cite{Schettler:2010wi}.

{According to the inflationary cosmology with baryogenesis and {DM}/energy, physics beyond the Standard model (BSM) underlying the basis elements of modern cosmology can provide many scenarios of entropy production (see, e.g., \cite{khlopovPPNP, Chaudhuri:2021ppr,Chaudhuri:2021rwt, Chaudhuri:2021vdi} for review and references). In this article we have considered the entropy production within SM and the minimal extension of SM, namely, two Higgs Doublet Model (2HDM) and discuss the effects of \mbox{this mechanism}.}


{According to the $\Lambda-$CDM (Lambda cold dark matter) model of the universe at the present time, the $\Omega-$parameter of the universe for the cold DM is $\Omega_c\simeq 0.265(7)$ i.e., around $27\%$ of the total matter-energy of the today's universe is added from DM sector. DM only interacts by gravitational force with the SM members and thus it makes the circumstances really complex for the cosmologists to claim with certainty about the characteristics and constituents of DM sector. Due to this reason, numerous models of DM have been proposed for the last a few decades. All those models can be categorized into three groups - cold dark matter, warm dark matter (see \cite{1306.2314}), and hot dark matter (see{ p.178}  
of~\cite{Bambi:2015mba}). There are several proposed members of these three groups. Among them the promising candidates are priomordial black holes~\cite{Carr,PBHrev}, Axion~\cite{2105.01406} or Fuzzy DM or Axion like particles~\cite{ALP}, Weakly interacting massive particles~\cite{1903.03026}, 4th generation DM~\cite{astro-ph/0511796}, neutrinos (both cold and hot DM possible)~\cite{Ibarra, Sidhu da}, fermionic DM~\cite{fermionic DM}, gravitino~\cite{gravitino} etc.  Howbeit, the calculation of this script is independent of the particular DM model as long as their production or freezing out or freezing in is not connected with the EWPT mechanisms we considered or affect their parameters and frozen out event has been completed before the beginning of EWPT.  If the DM are produced during or later stage (e.g., \cite{DM during ewpt}), the dilution will be change due to increase in number density of DM particles. The scenario of production of DM before EWPT is already well known in literature (example~\cite{ALP,Fluid}). But the temperature scale of their decoupling from radiation depends their mass and cross section of their reaction with the SM particles.

 In this article, entropy density implies the entropy of the relativistic particles (photons) which dominates the entropy of the
universe because there are significantly more photons than baryons and non-baryons in our universe. 
For non-relativistic matter, chemical potential $\mu_{\rm chm. pot.}$ is not negligible and conservation of entropy can be expressed as continuity equation of entropy current. 
If the dark matter is in the form of cold dark matter, its entropy density
behaves as non-relativistic. If it is in the form of hot dark matter, then entropy density of DM behaves as relativistic and follow Equation \eqref{ent-consv}. For warm dark
matter, which is an intermediate particle between hot and cold dark matter then it can behave
as both. 
Disquisition about the entropy of DMs and non-relativistic particles is beyond the scope of this work. 

}

The largest dilution of {DM} as a result of entropy influx during the course of universe expansion took place during EWPT. According to electroweak (EW) theory when $T>T_c$, where $T_c$ is the critical temperature of EWPT, $\expval{\phi}=0$ of the Higgs field, see~\cite{Bochkarev:1990fx} but for $T<T_c$, $\expval{\phi}$ increases to the second vacuum expectation value $\eta$. Such a nonequilibrium state violates the entropy conservation law releasing entropy into the primeval plasma.


In the SM, EWPT can be of second order or a smooth crossover. In this work, we have considered EWPT as a second order phase transition. The influx of entropy can significantly increase if EWPT is of the first order, which happens in multi Higgs extension of the standard model, like {two-Higgs doublet model (2HDM)}. 

For $T\gg T_c$, the universe was in thermal equilibrium, and relativistic particles dominated the universe. 
The 
contribution to the overall energy
 density of the universe from those who were already massive (e.g., decoupled DM) was also insignificant.
To this extend, the entropy density per unit comoving volume follows the conservation law, as mentioned in Equation \eqref{ent-consv}.
In this scenario, the sum of the energy and the pressure density can lead up to 
\begin{equation}
\label{eq:rho+P}
\rho_r +P_r \sim g_* T^4. 
\end{equation}
where $g_*$ is the effective number of particle species at or near the EWPT.  

It is to be noted that $g_*$ is not constant but varies over time. It depends on the components of the primordial plasma. Equations~\eqref{ent-consv} and~\eqref{eq:rho+P} imply:
\begin{equation}
T \sim a^{-1}.
\end{equation}

If at a later stage the thermal equilibrium is broken, 
$s$ and thus $g_* \left(T\right) a^3 T^3$ increases as entropy should always increase or remain 
unchanged. The biggest contribution comes from the heaviest particle whose mass ($m(T)$) is closest to the temperature $T$, i.e., $m(T)<T$. Here the product of the effective number of particles to the temperature is unchanged and the rise in entropy is dominated by the change in the scale factor, i.e., $a^3T^3$, suggesting the influx of entropy happening over $g_*a^3T^3$.
%
And hence the net entropy release is given by
\be
\frac{\delta s}{s} = \frac{(a_c \, T_c)^3 - (a \, T)^3}{(a_c\, T_c)^3}.
\ee

In this work, we show that the dilution { happens in EWPT in both SM and 2HDM for all the DM candidates, created, frozen in or frozen out before EWPT.}  The paper is organized as follows. In the next two Sections \ref{SM} and \ref{2HDM} we calculate the entropy influx due to SM and 2HDM into the primeval plasma and show the dilution of pre-EWPT {DM} density. These are followed by a generic conclusion { that though this effect cannot influence predicted DM density of such popular DM candidates as WIMPs or QCD axions, which are frozen out or formed after EWPT, the dilution of the pre-EWPT DM density of sterile neutrinos, superheavy gravitinos or Primordial Black hole (PBHs) (see, e.g., \cite{khlopovPPNP} for review and references) is significant and should be taken into account in the detailed analysis of the corresponding cosmological scenarios.}

\section{Entropy Production and Dilution of Frozen Out {DM} Density within the Framework of the SM} \label{SM}

The Lagrangian of the EWPT in SM framework containing only a single Higgs field is given as:
\be \label{lagsm}
\mathcal{L}=g^{\xi \tau} \partial_{\xi}\phi^{\dagger}\partial_{\tau}\phi - U_{\phi}(\phi)+\sum_j i \left[g^{\xi \tau} \bar{\chi_j}\gamma_{\xi} \partial_{\tau} \chi_j - U_j (\chi_j)\right] +\mathcal{L}_{int} 
\label{eq:SM Lagrangian}
\ee
where $\phi$ is the Higgs field with potential $U_{\phi}(\phi)$. $\chi_j$ represents the fermionic fields which gain masses during or immediately after EWPT, $U_j(\chi_j)$ is the potential of the field $\chi_j$, and the metric is flat FRW background with the signature of the metric $g_{\xi \tau}=(+,-,-,-)$ with the Greek indices in sub- or superscripts running from $0$ to $3$.  The interaction part of the above Lagrangian density is
\be \label{19}
\mathcal{L}_{int}=\phi \sum_j g_j \bar{\chi_j} {\chi}_j,
\ee
where $g_j$ are the Yukawa coupling constants, and the summation is over all fields $\chi_j$.

In Equation \eqref{eq:SM Lagrangian}, we have used simplistic Lagrangian density 
instead of the original form of the Lagrangian density described in~\cite{Logan:2014jla}. The reason behind this is that the detailed form of the Gauge bosons and fermionic sector, as we will see below, is not needed for our calculations, and the reader, thus, can get the main working principle at ease. However, for completeness, we have used the exact form of the Lagrangian density in Section \ref{2HDM} for the case of 2HDM.

The homogeneous and isotropic classical field $\phi(t)$ follows the equation of motion 
%
%
as
\begin{eqnarray}
&&\ddot \phi + 3\mathcal{H} \dot \phi + U'_\phi (\phi) -  \sum_\chi g_{j} \bar{\chi_j} \chi_j = 0,
\label{eqom1} 
\end{eqnarray}
where the dot represents derivative w.r.t. time $t$, $\mathcal{H} = \dot a /a $ is the Hubble parameter for the cosmological scale factor $a(t)$ and $U'_\phi = dU_\phi /d\phi$. 
.

We have considered the following SM 
potential 
\begin{eqnarray} 
\label{23}
U_\phi (\phi) = \frac{\lambda}{4}(\phi^2-\eta^2)^2+  \frac{T^2 \phi^2}{2}\,\sum_j \,h_j \left( \frac{m_j(T)}{T} \right) , 
\end{eqnarray}

Here, $T$ and $m_j(T)$ are the temperature of the relativistic fluid and mass of the $\chi_j$-particle at temperature $T$, respectively.

The last term depends on the temperature and arises from (\ref{19}) after thermal averaging and involves the contributions of $\phi$ itself and of all particles $\chi_j$. The summation is made over all particles and the function $h_j  (m_j /T) \propto g_j^2$ and is positive always. For $T > m_j (T)$ the function takes a noticeable value and it is negligible for $T < m_j(T)$.

According to the experiment, the vacuum expectation value (vev) of 
the Higgs field
$\phi$ is equal to $v_{\rm sm}\equiv\eta=246$ GeV and the quartic
self-coupling of $\phi$ is $\lambda=0.13$, see \cite{Melo}.

Within the framework of SM the fermionic contribution is important and their masses at zero temperature is given by $m_f = g_f \eta$, $g_f$ is the Yukawa coupling between the Higgs and fermionic members. The thermal masses of the fermions are proportional to the temperature which depends on the value of $\phi$ at the minima of the potential (\ref{23}).

\begin{equation}
 \phi^2_{min} (T) = \eta^2 - (T^2/\lambda) \sum_j h_j \left(\frac{m_j (T)} {T}\right) 
\label{24}
\end{equation}
and 
analogously
\begin{equation}
 m^2_f (T) = g^2_f \phi^2_{min} (T) = g^2_f \left[ \eta^2 - (T^2/\lambda) \sum_j h_j \left(\frac{m_j (T)} {T}\right) \right]  .
\label{25}
\end{equation}

The dilution of preexisting baryon asymmetry and the dilution of {DM} density can be computed from Equation (\ref{ent-consv}) and to do this we need $\mathcal{P}$ and $\rho$ which can be obtained from the energy--momentum tensor: 
\begin{eqnarray}\label{32a}
T_{\mu\nu} &=& \partial_\mu \phi \, \partial_\nu \phi - 
g_{\mu\nu} \left( g^{\alpha\beta} \partial_\alpha\phi \, \partial_\beta \phi - U_\phi (\phi) \right) 
\\ \nonumber 
&+& \sum_j \left[ \partial_\mu\bar{\chi_j} \, \partial_\nu\chi_j + \partial_\nu\bar{\chi_{j}} \, \partial_\mu \chi_{j} -
g_{\mu\nu} \left( g^{\alpha\beta} \partial_\alpha\bar{\chi_{j}} \, \partial_\beta\chi_{j} - U_j (\chi_j) + 2  \mathcal{L}_{int}  \right) \right] ,
\end{eqnarray}
where $\mathcal{L}_{int}$ is already specified in Equation (\ref{19}).

The energy and pressure density can be calculated from Equation (\ref{32a}) and is given by:
\begin{eqnarray}
\rho &=& \dot\phi^2/2 + U_\phi (\phi)  + 
\sum_j \left[ \dot {\bar{\chi_j}} \dot \chi_j + \partial_l\bar{\chi_{j}}\, \partial_l\chi_{j }/a^2  + U_j (\chi_j) \right] - \mathcal{L}_{int}; \label{33} \\
\mathcal{P} &=& \dot\phi^2/2 - U_\phi (\phi)  + 
\sum_j \left[ \dot {\bar{\chi_j}} \dot \chi_j - (1/3)\partial_l\bar{\chi_{j}}\, \partial_l \chi_{j} /a^2  - U_j (\chi_j) \right] + \mathcal{L}_{int},
\label{34} 
\end{eqnarray}
where $\partial_i \chi$ is 
the 
derivative with respect to spatial coordinate.

To simplify the calculation, it is assumed that the energy density is the summation of the energy density of $\phi (t)$ which sits at the minimum of the potential 
and relativistic particles, hence $\rho$ becomes:
\begin{equation}
\rho \approx U_\phi (\phi_{min} ) + \frac{\dot \phi^2}{2}  + \frac{\pi^2 g_*}{30} T^4,
\label{37}
\end{equation}
along with
\begin{equation}
{\mathcal P} + \rho  \approx \dot \phi^2 + \frac{4}{3}\,\frac{\pi^2 g_*}{30} T^4,
\label{38}
\end{equation}

At or around the electroweak phase transition, $g_*\sim 10^2$ is the effective number of relativistic particles. It is a temperature-dependent function that decreases as the cosmological cooling process progresses. In the limit of instant thermalization, Equation (\ref{37})  holds true.


Under this assumption, the oscillations of $\phi$ around $\phi_{min} $ follows fast damping, thus we take $\dot \phi = \dot \phi_{min}$ and neglect higher order terms as the evolution of $\phi_{min}$ is induced by the universe expansion which is very slow. A single differential equation of the temperature (or scale factor) is thus obtained.
Finally the equation of energy conservation can be \mbox{rewritten as}
\begin{equation}
\frac{\dot T}{T} \left[ 
h_{tot} (m) \eta^2  T^2 \left( 1  - \frac{T^2}{T^2_c}  \frac{ h_{tot} (m) }{ h_{tot}(0) } \right)  + \frac{4 \pi^2 g_*}{30} T^4\right]  
= - 4 \mathcal{H} \frac{\pi^2 g_*}{30} T^4 .
\label{39}
\end{equation}
where $\sum h_j (m_j/T) \equiv h_{tot} (m)$ and  $\sum h_j (0) \equiv h_{tot} (0) $.

\begin{figure}[h!]
\includegraphics[]{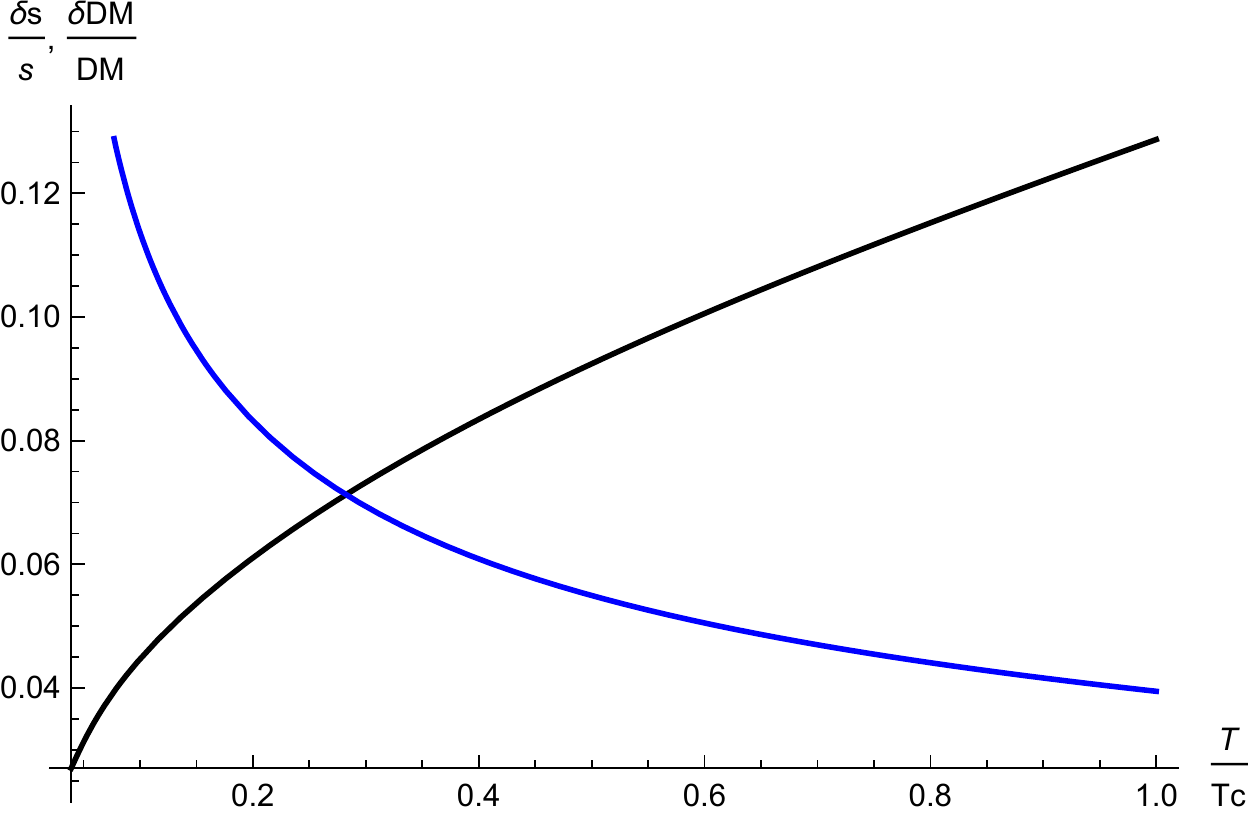}
\caption{The black line shows the entropy released into the plasma, and the blue line shows the corresponding dilution of {DM} density. The net dilution is $\sim$13\%.}
\label{rat-F2-FLN}
\end{figure}

Following the above assumptions, the relative influx of entropy which leads to the corresponding dilution of {DM} is given by:
\begin{equation}
\frac{\delta s}{s} = \sum_f x_{j,min}^{6\nu_f} \exp\left[ 3\nu_f \left(\frac{1}{x_{f,min}^2} -1 \right)\right]  - 1 ,
\label{46}
\end{equation}
where $x = T/T_c$, 
$\nu_f=\frac{\kappa \eta ^2}{ 2 T_c^2}$ and $\kappa= \frac{30 h_{tot}(m)}{4 \pi^2 g_*}$. 
As a result of entropy inflow into the primordial plasma, the total dilution of frozen-out DM particles is shown in {the
Figure \ref{rat-F2-FLN} 
}.

Within the framework of the SM, the EWPT is assumed to be of second order. Even though the dilution is quite small, it is noticeable. 
The $t$-quark {(top-quark)}, the heaviest SM particle contributes mostly to the influx of the entropy.
In the extended versions of SM with multi Higgs, the influx and in return the dilution is quite larger as seen in Section \ref{2HDM}.

\section{Entropy Production and Dilution of Frozen Out {DM} Density within the Framework of 2HDM} \label{2HDM}
In this section, we explore entropy generation and the dilution of frozen out DM density within the context of the two-Higgs doublet model (2HDM).
%
It is the most popular extension of the scalar sector of the SM. An additional scalar doublet is necessary for the supersymmetric models. Non-supersymmetric of this model includes a successful electroweak baryogenesis \cite{EWBG}. There are various sectors of 2HDM like Type-I 2HDM, Type-II 2HDM, etc. A detailed review of 2HDM describing all the sectors can be found \mbox{in \cite{Branco:2011iw}.} In this work, we confine ourselves to real Type-I 2HDM where the charged fermions only couple to second doublet.

The Lagrangian for the 2HDM scenario is given in the following form:
\begin{equation} 
\mathcal{L}= \mathcal{L}_f  + \mathcal{L}_{\rm Yuk} + \mathcal{L}_{\rm gauge, kin}  + \mathcal{L}_{\rm Higgs}. \label{Eq: Total lagrangian}
\end{equation}
and each term on the R.H.S. is defined as follows:
The first term is the kinetic term originating due to the fermionic fields:
\begin{eqnarray}
\mathcal{L}_f &=&
 \sum_{j}i (\bar{\Psi}^{(j)}_L \slashed{D}\Psi^{(j)}_L + \bar{\Psi}^{(j)}_R \slashed{D}\Psi^{(j)}_R  ) \\ &=&i\bar{\Psi}_L \gamma^\xi (\partial_\mu + ig W_\mu + i g'Y_L B_\mu) \Psi_L   \nonumber \\&& + i\bar{\Psi}_R  \gamma^\mu (\partial_\mu + ig W_\mu + i g'Y_R B_\mu) \Psi_R.     \label{Eq: Fermionic Lagrangian}
\end{eqnarray}

Here, $L$ and $R$ are the left and right chiral field 
of that fermion, $\slashed{D}$ is the
the gauge covariant derivative in Feynman notation
(Feynman notation is defined as $\slashed{A}=\gamma^{\mu} A_\mu$, where $\gamma$ are the gamma matrices) and $j$ runs over all fermionic species (the field $\Psi_j$). In addition, $g$ is the coupling constant. The partial derivative follows from the previous section, e.g., $\partial _0 = d/dt$ and $\mu$ runs from $0$ to $3$,~\cite{Karmakar,Karmakar:2020mds}.

The second term of Equation~(\ref{Eq: Total lagrangian}), the Yukawa interaction term is given by  
\be
\mathcal{L}_{\rm Yuk}= -\left[ y_e \bar{e_R} \Phi_a^\dagger L_L + y_e^* \bar{L_L} \Phi_a^\dagger e_R  + \cdots \right],
\ee
where $y_e$ is a dimensionless constant which is complex in nature, $\Phi_a$ ($a=1,2$) is a $SU(2)_L$ doublet, and since the Lagrangian needs to be gauge invariant, it is coupled with another $SU(2)_L$ fermion $L_L$. $e_R$ follows from above, is the right chiral electron field, and so are the other fermions, e.g., quarks, neutrinos, etc. 
\par The third term $\mathcal{L}_{\rm gauge, \,kin} $ represents $U(1)$ invariant  kinetic term of four gauge bosons ($W^i, \, i=1,2,3$, and $B$). It can be written in the following manner:
\begin{eqnarray}
\mathcal{L}_{\rm gauge,\, kin} = -\frac{1}{4}G^i_{\mu \nu}{G^i}^{\mu \nu}-\frac{1}{4}F^B_{\mu \nu}{F^B}^{\mu \nu},          \label{Gauge-kinetic Lagrangian}
\end{eqnarray}
where $G^i_{\mu \nu}=\partial_\mu W^i_\nu-\partial_\nu W^i_\mu - g \epsilon^{ijk}W_\mu^j W_\nu^k$ and $F^B_{\mu \nu}=\partial_\mu B_\nu-\partial_\nu B_\mu$ 
with $\epsilon_{ijk}$ is the Levi-Civita symbol. 

The fourth term is the Lagrangian for the Higgs boson doublets: 
\vspace{-9pt}
\begin{eqnarray}
\mathcal{L}_{\rm Higgs} &=& (D^\mu \Phi_1)^\dagger (D_\mu \Phi_2) + (D^\mu \Phi_1)^\dagger (D_\mu \Phi_2) -V_{\rm tot}(\Phi_1,\Phi_2)  \nonumber \\  
&=& \{(\partial_\mu  + ig T^i W^i_\mu + i g'Y B_\mu ) \Phi_1  \}^\dagger\{(\partial_\mu +  ig T^i W^i_\mu + i g'Y B_\mu ) \Phi_1  \}    \\
&&+ \{(\partial_\mu  + ig T^i W^i_\mu + i g'Y B_\mu ) \Phi_2 \}^\dagger\{(\partial_\mu +  ig T^i W^i_\mu + i g'Y B_\mu ) \Phi_2 \}    - V_{\rm tot}(\Phi_1,\Phi_2, T).   \nonumber  \label{Eq:eq1}
\end{eqnarray} 

For simplification, we define $\mathcal{W}_\mu=gT^iW^i_\mu + g'Y B_\mu$, we've from Equation (\ref{Eq:eq1}): 
\begin{eqnarray}
	 \mathcal{L}_{\rm Higgs,\, kin}&=&(\partial^\mu {\Phi_a}^\dagger )(\partial_\mu \Phi_a) - i ({\mathcal{W}^\mu}\Phi_a)^\dagger (\partial_\mu \Phi_a) + \\ \nonumber
	 && i (\partial^\mu{\Phi_a}^\dagger)\mathcal{W}_\mu\Phi_a+  (\mathcal{W}^\mu\Phi_a)^\dagger \mathcal{W}_\mu\Phi_a.
\end{eqnarray}

The type-I 2HDM potential and correction terms are as follows:
\begin{equation} \label{2HDM potential}
    V(\Phi_1,\Phi_2,T)=V_{tree}(\Phi_1,\Phi_2)+V_{CW}(\Phi_1,\Phi_2)+V_{T}(T) {+ V_{\rm daisy} (T)},
\end{equation}

The tree level potential is given by:
\begingroup\makeatletter\def\f@size{10}\check@mathfonts
\def\maketag@@@#1{\hbox{\m@th\normalsize\normalfont#1}}%
\begin{align} 
V_{\rm tree}(\Phi_1,\Phi_2)   =& m_{11}^2 \Phi_1^\dagger \Phi_1 + m_{22}^2 \Phi_2^\dagger \Phi_2 - \left[m_{12}^2 \Phi_1^\dagger \Phi_2 + m_{12}^* \Phi_2^\dagger \Phi_1 \right] + \frac{1}{2} \lambda_1 \left(\Phi_1^\dagger \Phi_1\right)^2 \nonumber \\ 
&+  \frac{1}{2} \lambda_2 \left(\Phi_2^\dagger \Phi_2\right)^2  + \lambda_3 \left(\Phi_1^\dagger \Phi_1\right)\left(\Phi_2^\dagger \Phi_2\right) + \lambda_4 \left(\Phi_1^\dagger \Phi_2\right)\left(\Phi_2^\dagger \Phi_1\right) \\
&+ \left[\frac{1}{2} \lambda_5  \left(\Phi_1^\dagger \Phi_2\right)^2 + \frac{1}{2} \lambda_5^*  \left(\Phi_2^\dagger \Phi_1\right)^2 \right]\nonumber .
\end{align}
\endgroup

Here, $m_{12}$ is a function of the mixing angle $\beta$ which follows $tan\beta=v2/v1$, where $v_1$ and $v_2$ are the expectation values of $\Phi_1$ and $\Phi_2$, respectively.
Despite the fact that $\tan\beta$ is a physical parameter, it is still feasible to redefine the two doublets and arrive at a basis where the full vev is totally contained within one of the two doublets, which is named as Higgs basis, see ~\cite{Bernon:2017jgv}.
%
%
The quartic coupling constants are real for our scenario and hence $\lambda_5^*=\lambda_5$. The explicit expressions for the couplings are given by:
\begin{align}
    \lambda_1 &= \frac{1}{v_{\rm sm}^2 \cos^2 \beta}  \lt( - \mu^2 \tan\beta + m_h^2 \sin^2 \alpha + m_H^2 \cos^2\alpha\rt) \\
    \lambda_2 &= \frac{1}{v_{\rm sm}^2 \sin^2 \beta} \lt( - \frac{\mu^2}{\tan\beta} + m_h^2 \cos^2 \alpha + m_H^2 \sin^2\alpha\rt)   \\
    \lambda_3 &= \frac{1}{v_{\rm sm}^2 }  \lt( - \frac{2 \mu^2}{\sin 2\beta} + 2 m_{H_\pm}^2 + \lt(m_H^2 - m_h^2 \rt)\frac{\sin 2 \alpha}{\sin 2 \beta}\rt) \\
    \lambda_4 &= \frac{1}{v_{\rm sm}^2 }  \lt(  \frac{2 \mu^2}{\sin 2\beta} + m_A^2 - 2 m_{H_\pm}^2 \rt) \\
    \lambda_5 &= \frac{1}{v_{\rm sm}^2 }  \lt( \frac{2 \mu^2}{\sin 2\beta} - m_A^2  \rt)
\end{align}

All the masses (i.e., $m_h, m_{H_\pm}, m_H, m_A$ and other particles) are dependent on the second minimum of the potential. For detailed calculation of the masses, see~\cite{Chaudhuri:2021agl}. Here, $\alpha$.

The other terms in Equation \eqref{2HDM potential} are given below:
\begin{align}
&V_{\rm CW}\left(v_1+v_2 \right)=\sum_j \frac{n_j}{64\pi^2} (-1)^{2s_j}m_j^4\left(v_1,v_2\right)\left[ \log\left( \frac{m_j^2 \left(v_1,v_2 \right)}{\mu^2} \right) - c_j \right] \\
&V_T=\frac{T^4}{2\pi^2}\left( \sum_{j={\rm bosons}} n_j J_{B}\left[\frac{m_j^2(v_1,v_2)}{T^2}\right]  + \sum_{j={\rm fermions}} n_j J_{F}\left[\frac{m_j^2(v_1,v_2)}{T^2}    \right] \right)
\end{align}
where $s_j$ is the spin of the $j-$th particle, and $\mu$ is the  renormalisation scale, whose value is takes as $246$~GeV. Detailed description of the parameters are done in \cite{Chaudhuri:2021agl}.

$J_B$ and $J_F$ are approximated in series expansion at high temperature in the Landau gauge up to the leading orders:
\begin{eqnarray}
T^4 J_B \left[\frac{m^2}{T} \right] &&= -\frac{\pi^4 T^4}{45} + \frac{\pi^2}{12}T^2 m^2 - \frac{\pi}{6}T (m^2)^{3/2} - \frac{1}{32}m^4 \ln \frac{m^2}{a_b T^2} + \cdots,   \\
T^4 J_F \left[\frac{m^2}{T} \right] &&=  \frac{7\pi^4 T^4}{360} - \frac{\pi^2}{24}T^2 m^2 - \frac{1}{32}m^4 \ln \frac{m^2}{a_f T^2} + \cdots, 
\end{eqnarray}
where $a_b=16a_f=16\pi^2 \exp(3/2-2\gamma_E)$ with $\gamma_E$ being the Euler--Mascheroni constant.

The daisy correction term is given by:
\be
V_{\text{daisy}}(T)=-\frac{T}{12\pi} \left[\sum_{j=1}^{n_{\text{Higgs}}}
\left(\bar{(m}^2_j)^{3/2}  - (m_j^2)^{3/2}\right) +\sum_{j=1}^{n_{\text{gauge}}}
\left(\bar{(m}^2_j)^{3/2} - (m_j^2)^{3/2} 
\right) \right]
\label{eq:AE2} 
\ee

As the universe cools down, a second local minimum appears. As the temperature of the plasma drops to $T_c$ the second local minimum takes the form $(\left<\Phi_1\right>=v_1,\left<\Phi_2\right>=v_2)$ and become degenerate with the global one at $(\left<\Phi_1\right>=\left<\Phi_2\right>=0)$. Thus the critical temperature is obtained by the following expression:
 \begin{equation}
     V_{\rm tot}\left(\Phi_1=0,\Phi_2=0,T_c \right)= V_{\rm tot}\left(\Phi_1=v_1,\Phi_2=v_2,T_c \right).
 \end{equation}

The energy density of the homogeneous classical field $\Phi$: 

\begin{align}
	\rho =&\partial^0 \Phi_a^\dagger \partial^0\Phi_a - (\mathcal{W}^0 \Phi_a)^\dagger \mathcal{W}^0\Phi_a 
	- (\mathcal{W}^j \Phi_a)^\dagger \mathcal{W}_j\Phi_a  \nonumber \\
	&+\lt[V_{\rm tot}(\Phi_1,\Phi_2, T)  - \mathcal{L}_{\rm gauge, kin} - \mathcal{L}_f  - \mathcal{L}_{\rm Yuk}  
	\rt]     \label{Eq: Expression for energy desnity}
\end{align}
\noindent where,
the spatial derivatives of Higgs fields vanish due to the condition of homogeneity and isotropy.
Similarly,
\begin{eqnarray}
\rho + P= 2\partial^0\Phi_a \partial^0 \Phi_a^\dagger - i ({\mathcal{W}^0}\Phi_a)^\dagger (\partial_0 \Phi_a) + i (\partial^0{\Phi_a}^\dagger)\mathcal{W}_0\Phi_a.
\label{Faltu equation 5}
\end{eqnarray}

Following the assumption from Section \ref{SM} (before Equation \eqref{39}), the oscillations follow fast damping and hence higher order terms are neglected.
And hence we simplify \mbox{Equation \eqref{Faltu equation 5}} to the following expression:
\be \label{22}
\rho =\Dot{\Phi}_{a, {\rm min}}^2 
+  V_{\rm tot}(\Phi_1,\Phi_2, T) + \frac{g_* \pi^2}{30} T^4.
\ee

The endmost term on the right hand side of Equation (\ref{22}) is the energy density of the relativistic particles that have not obtained mass until EWPT. The Yukawa interaction between fermions and Higgs bosons, as well as the energy density of fermions, gauge bosons, and the Higgs-gauge boson interaction contribute to this term.

Entropy is conserved, and the density of the frozen out {DM} species remains constant as long as the primeval plasma, which is assumed to be an ideal fluid, remains constant. But below the critical temperature $T_c$, the universe goes into the state of non-equilibrium. This situation is rather useful since 2HDM can successfully explain electroweak baryogenesis and deviation from thermal equilibrium is one of the primary conditions for baryogenesis. Due to this non-equilibrium state, the entropy conservation law breaks and the frozen out {DM} density starts diluting due to the influx of entropy into the plasma.

In order to determine the dilution of the frozen out species it is important to solve the evolution equation for the conservation of the energy density,
\begin{equation} \label{fried}
\Dot{\rho}=-3\mathcal{H}(\rho+P).  
\end{equation}

Unlike the previous section, it is not easy to proceed forward with analytical calculations. Hence numerical calculations are done from henceforth. {For this purpose we use the public code: Beyond the Standard Model Phase Transitions (BSMPT) \cite{Basler:2020nrq,Basler:2018cwe}, 
 which is} 
which is written in C++ programming language and it can be employed for calculating the strength of the electroweak phase transition in extended Higgs sectors. This tool takes care the corrected terms of the tree-level potential, including loop-correction and daisy re-summation of the bosonic masses.

The analytical calculations for ${\delta s}/{s}$ were possible in the previous section because primarily because of the single Higgs field and the simplified Lagrangian that was used. However, in the case of 2HDM, there are two Higgs fields, and the situation gets more complex as we are taking the exact Lagrangian, and hence numerical calculations \mbox{are needed.}

As we have seen in the previous section, near $T_c$ the Higgs potential gets a degenerate minimum
and as the temperature falls off to the mass of any relativistic component, that component acquires mass, and becomes non-relativistic.
Assuming this process is instantaneous and there is a change in $g_*$, entropy density $s$ increases. With the evolution of scale factor (and with time actually), $T$ changes, and with the temperature the value of $\phi_{\rm min}$ where the potential gets second minima. The variation of  $\phi_{\rm min}$ with $T$ can be obtained from BSMPT, as we will see below. To find the variation of $T$ with $a$, we need Equation \eqref{fried}.

\begin{figure}[h]
  \centering
  \begin{minipage}[b]{0.3\textwidth}
    \includegraphics[width=\textwidth]{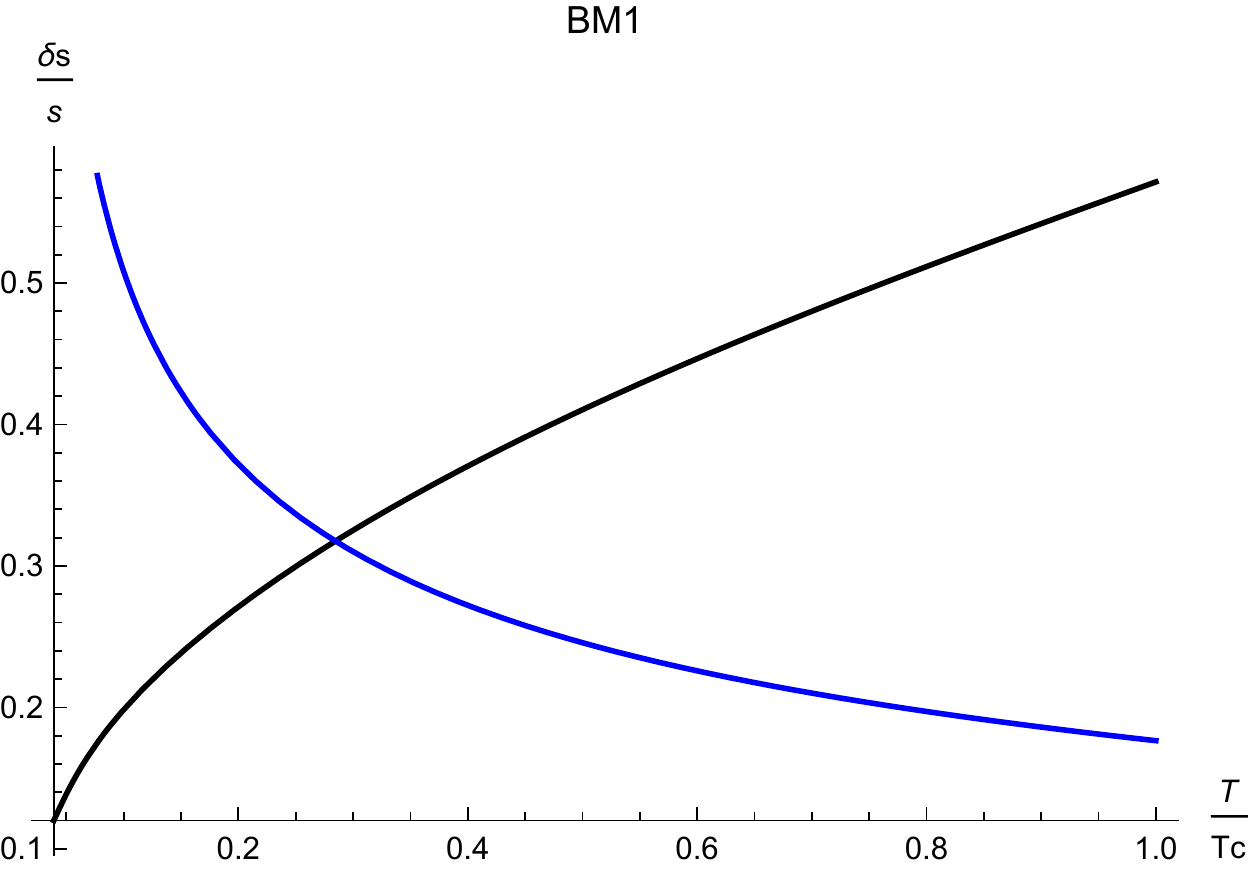}
  \end{minipage}
  \hspace*{.1cm}
  \begin{minipage}[b]{0.3\textwidth}
    \includegraphics[width=\textwidth]{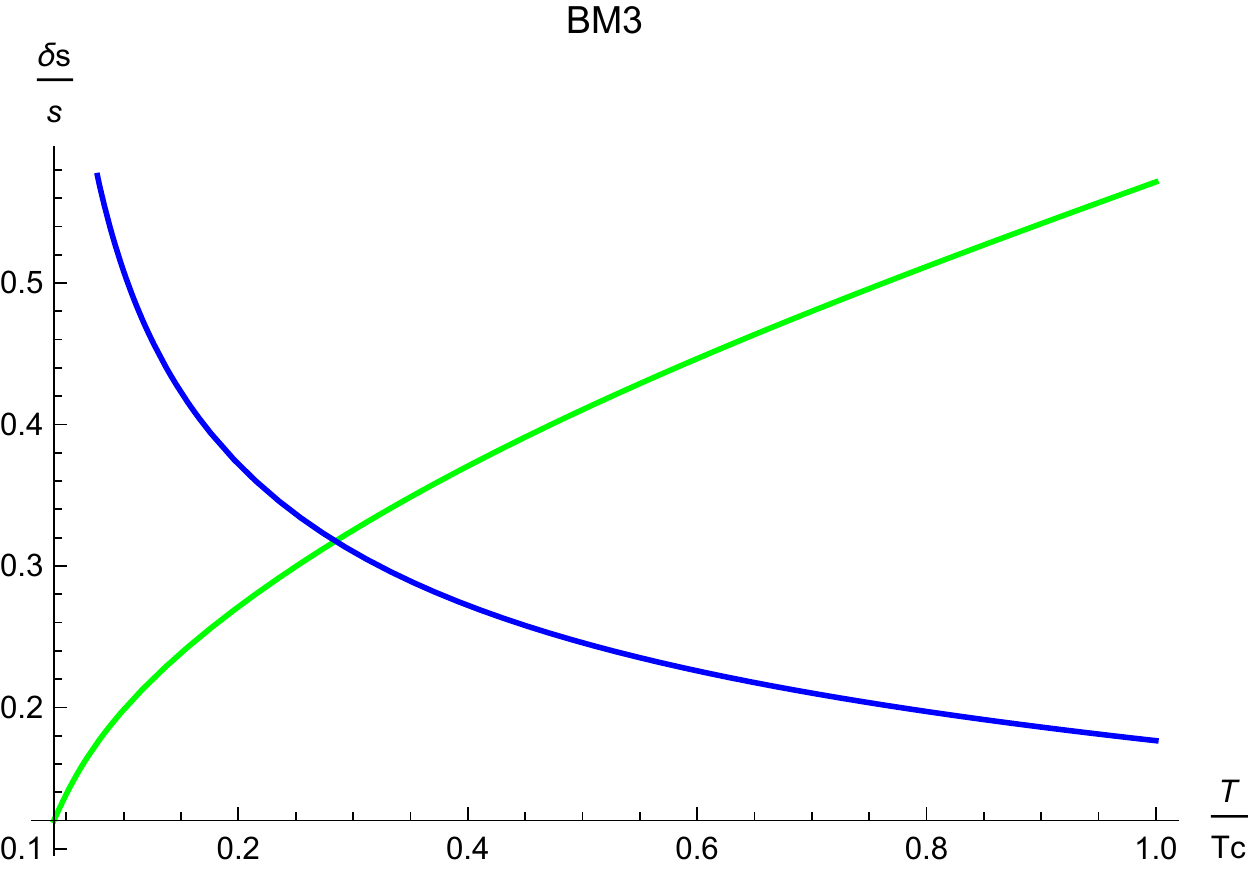}
      \end{minipage}
    \hspace*{.1cm}
    \begin{minipage}[b]{0.3\textwidth}
    \includegraphics[width=\textwidth]{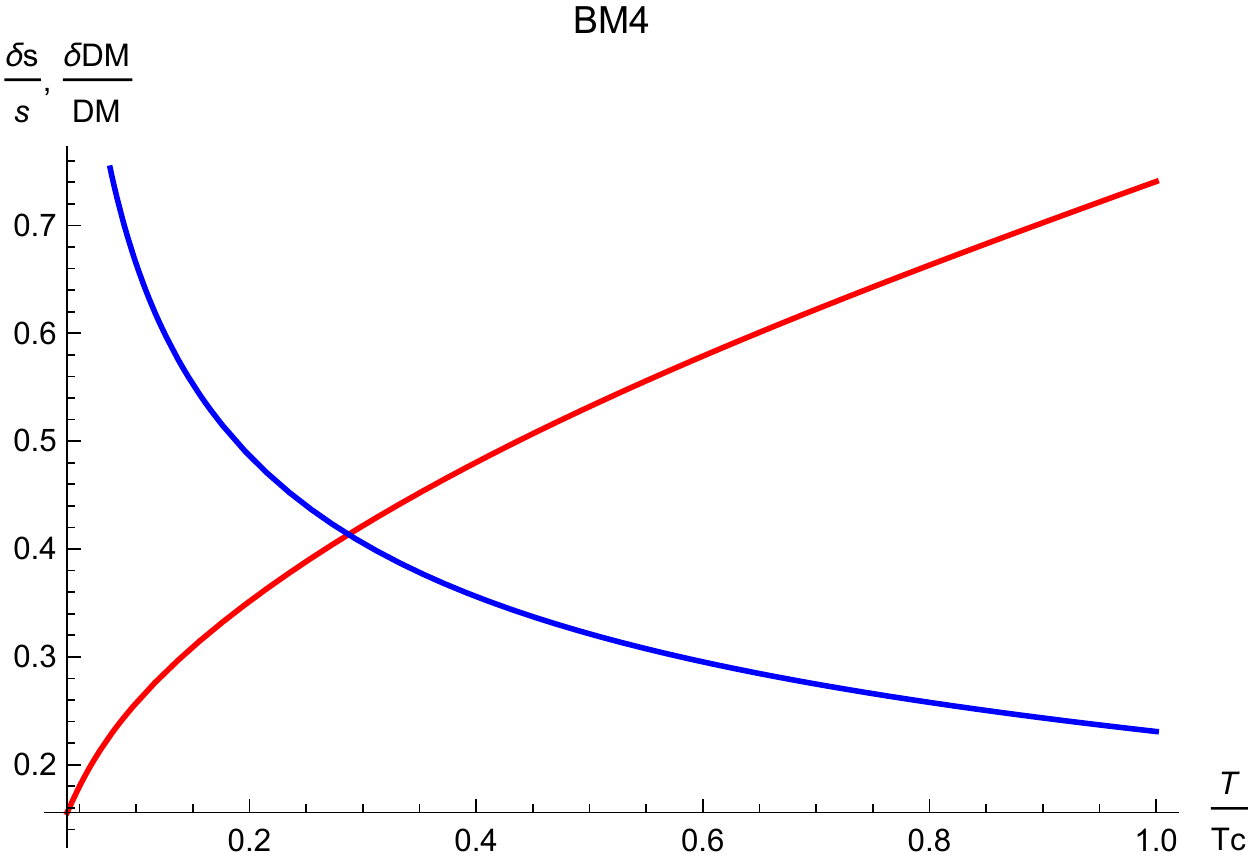}
      \end{minipage}
      \hspace*{.1cm}
  \begin{minipage}[b]{0.3\textwidth}
    \includegraphics[width=\textwidth]{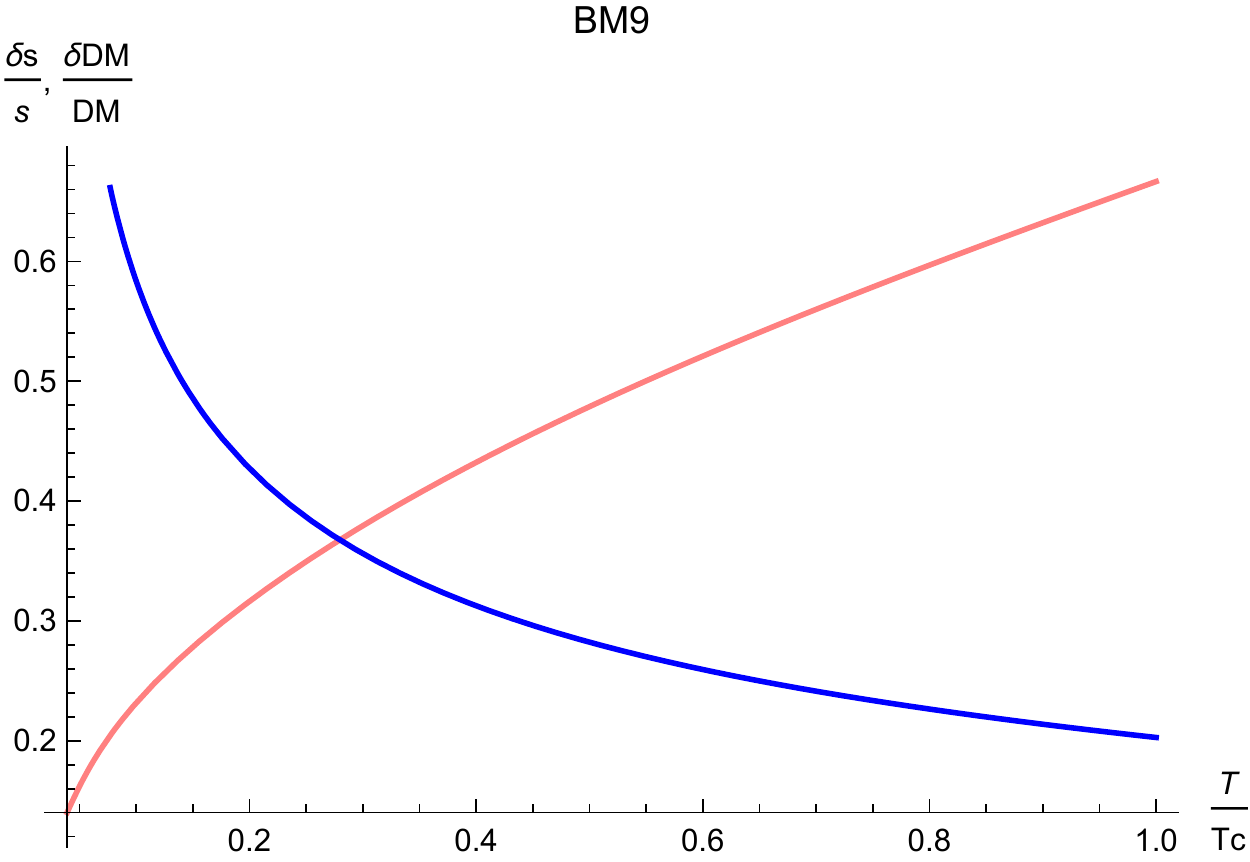}
      \end{minipage}
      \hspace*{.1cm}
  \begin{minipage}[b]{0.3\textwidth}
    \includegraphics[width=\textwidth]{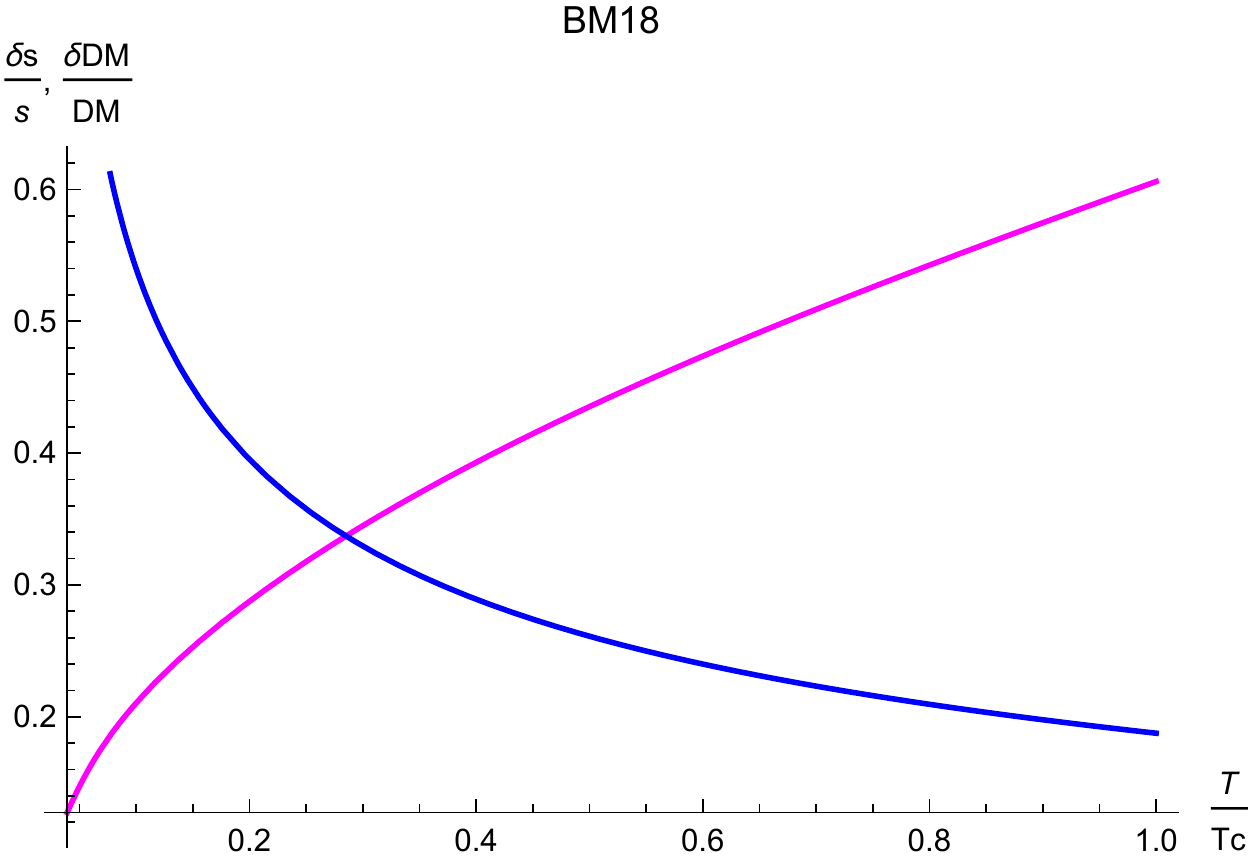}
      \end{minipage}
      \hspace*{.1cm}
  \begin{minipage}[b]{0.3\textwidth}
    \includegraphics[width=\textwidth]{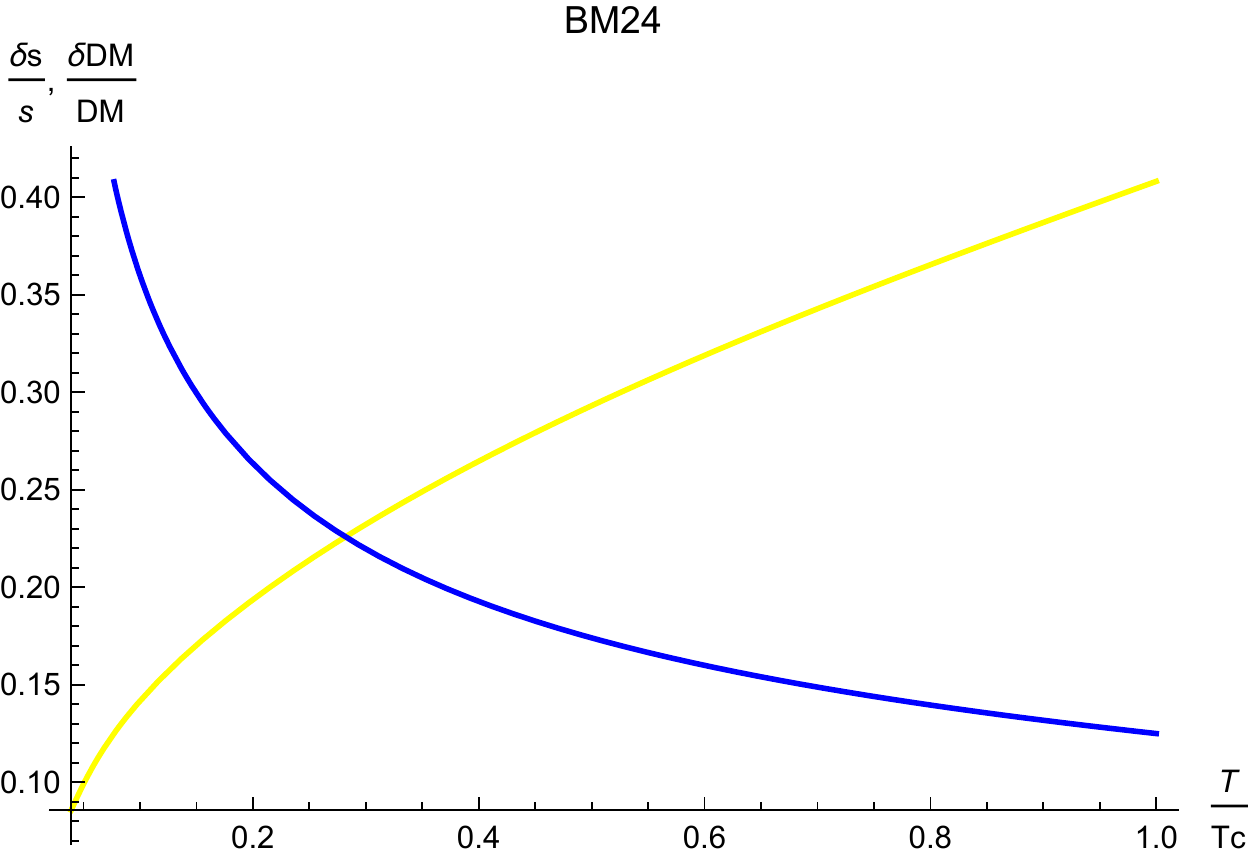}
      \end{minipage}
  \caption{For $6$ benchmark points the dilution factor,i.e., the amount of dilution of the frozen out {DM} particle density (blue lines) and the entropy influx factor (other coloured lines) are shown. Each plot corresponds to a particular benchamrk point which is mentioned within the plot itself. }
 \label{2hdm-entropy-fig}
 \end{figure}

The program is used to calculate the temperature dependent vev ($v$) and $T_c$. Here, we considered only the real sector of Type-I CP-conserving 2HDM potential and numerically computed the vevs, $T_c$ and the effective value of the potential $V_{eff}(T)$ for all the benchmark points in Table \ref{Table:1 Benchmark values} with the constraining condition $vev/T_c>0.01$. The differential \mbox{Equation \eqref{fried}} was computed numerically for all the benchmark points in Table \ref{Table:1 Benchmark values}, and the dilution factor, as well entropy production 
for some benchmark points, are shown in \mbox{Figure \ref{2hdm-entropy-fig}.}


\begin{table}
\footnotesize 

\centering
		\caption{2HDM Benchmark points for entropy production.}\label{Table:1 Benchmark values}
		\resizebox{\textwidth}{!}{
\begin{tabular}{cccccccccccccccc} 
			\toprule
			& \boldmath{$m_h$} \textbf{[GeV]}& \textbf{\boldmath{$m_H$} [GeV]} & \textbf{\boldmath{$m_{H^{\pm}}$} [GeV]}& \textbf{\boldmath{$m_A$} [GeV]} & \boldmath{$\tan\beta$} & \boldmath{$\cos \left(\beta-\alpha \right)$} & \textbf{\boldmath{$m_{12}^2~\text{GeV}^2$}} & \boldmath{$\lambda_1$} & \boldmath{$\lambda_2$} & \boldmath{$\lambda_3$} & \boldmath{$\lambda_4$} & \boldmath{$\lambda_5$} & \boldmath{$T_c$} & \boldmath{$vev/T_c$} & \boldmath{$\delta s/s [\%]$}\\
			\midrule
BM1 & $125$ & $500$ & $500$ & $500$ & $2$ & $0$ & $10^5$ & $0.258$ & $0.258$ & $0.258$ & $0$ & $0$ & $ 161.36$ & 1.4 & $57$ \\
	\midrule
BM2 &{"}
{$^\ddagger$} 
 &" &" & " &" & $0.06$ & " &$1.14$ & $0.037$ & $0.63$ & $0$ & $0$ & $ 167.95$ & 1.6 & $59$ \\
				\midrule
BM3 & " & " & " & " & $10$ & $0$ & 24,752.5 & $0.258$ & $0.258$ & $0.258$ &$0$ & $0$ &  $161.02$ & 1.4 & $56$\\
	\midrule
 BM4 & "& " & " & " & $10$ & $0.1$ & 24,752.5 & $4.13$ & $0.22$ & $4.15$ & $0$ & $0$ & $255.71$ & 1.9 & $73$\\
	\midrule
BM5 & " & " & " & $485$ & $2$ & $0.0$ & $10^5$ & $0.26$ & $0.26$ & $0.26$ & $-0.244$ & $0.244$ & $161.53 $  & 1.4 & $57$\\
	\midrule
BM6 & " & " & " & $485$ & $2$ & $0.07$ & $10^5$ & $1.28$ & $0.002$ & $0.7$ & $-0.244$ & $0.244$ & $169.81 $  & 1.7 & $60$\\
	\midrule
BM7 & " & " & " & $477$ & $2$ & $0.07$ & $10^5$ & $1.28$ & $0.002$ & $0.7$ & $-0.37$ & $0.37$ &  $169.53 $ & 1.7 & $60$ \\
	\midrule
BM8 & " & " & " & $485$ &  $10$ & $0.0$ & 24,752.5 & $0.258$ & $0.258$ & $0.258$ & $-0.244$ & $0.244$ & $160.76 $ & 1.3 & $56$ \\
	\midrule
 BM9 & " & " & " & $350$ & $10$ & $0.1$ & 24,752.5 & $4.13$ & $0.22$ & $4.15$ & $-2.1$ & $2.1$ & $209.87 $ & 1.8 & $68$ \\
	\midrule
BM10 & " & " & $485$ & $500$ & $2$ & $0.00$ & $10^5$ & $0.258$ & $0.258$ & $-0.23$ & $0.49$ & $0$ & $153.27 $ & 1.25 & $53$ \\
	\midrule
BM11 & " & " & $485$ & $500$ & $2$ & $0.07$ & $10^5$ & $1.28$ & $0.002$ & $0.21$ & $0.49$ & $0$ & $169.28 $  & 1.7 & $60$ \\
	\midrule
BM12 & " & " & $485$ & $500$ & $10$ & $0.0$ & 24,752.5 & $0.258$ & $0.258$ & $-0.23$ & $0.49$ & $0$ & $160.51 $  & 1.3 & $56$ \\
	\midrule
 BM13 & " & " & $485$ & $500$ & $10$ & $0.1$ & 24,752.5 & $4.13$ & $0.22$ & $3.66$ & $0.49$ & $0$ & $241.75 $  & 1.88 & $70$ \\
	\midrule
BM14 & " & " & $485$ & $485$ & $2$ & $0$ & $10^5$ & $0.258$ & $0.258$ & $-0.23$ & $0.24$ & $0.24$ & ${ 159.76 }$  & 1.3 & $59$ \\
	\midrule
BM15 & " & " & $485$ & $485$ & $2$ & $0.07$& $10^5$ & $1.28$ & $0.002$ & $0.21$ & $0.244$ & $0.244$  & ${ 168.61}$ & 1.7 & $59$ \\
\midrule
 BM16 & " & " & $485$ & $485$ & $10$ & $0$ & 24,752.5 & $0.258$ & $0.258$ & $-0.23$ & $0.244$ & $0.244$ &${ 160.19 }$  & 1.3 & $56$\\
	\midrule
BM17 &" & $485$ & $485$ & $485$ & $2$ & $0.0$ &94,090 & $0.258$ & $0.258$ & $0.258$ & $0$ & $0$ &${ 161.31 }$  & 1.4 & $57$ \\
	\midrule
BM18 & " & $485$ & $485$ & $485$ & $2$ & $0.07$ & 94,090 & $1.22$ & $0.02$ & $0.67$ & $0$ & $0$ & ${ 169.7 }$  & 1.7 & $60$ \\
	\midrule
 BM19 & " &  $485$ & $485$ & $485$  & $10$ & $0$ & 23,289.6 & $0.258$ & $0.258$ & $0.258$ & $0$ & $0$ & ${ 160.96}$  &1.3 & $57$  \\
	\midrule
 BM20 & " &  $485$ & $485$ & $485$ & $10$ & $0.1$& 23,289.6 & $3.9$ & $0.22$ & $3.9$ & $0$ & $0$  & ${ 230.18 }$  & 1.86 & $70$ \\
	\midrule
 BM21 & " &  $485$ & $485$ & $500$ & $2$ & $0$ & 94,090 & $0.258$ & $0.258$ & $0.258$ & $0$ & $0$ & ${ 161.31}$ & 1.4  & $57$  \\
	\midrule
 BM22 & " &  $90$ & $200$ & $300$ & $2$ & $0$ & $3240$ & $0.258$ & $0.258$ & $1.31$ & $0.3$ & $-1.35$ & ${ 150.76 }$ & 1.2 & $51$  \\
	\midrule
  BM23 & " &  $90$ & $200$ & $300$ & $10$ & $0$ & $801.98$ & $0.258$ & $0.258$ & $1.31$ & $0.3$ & $-1.35$ & ${ 135.38  }$ & 1.06 & $37$  \\
	\midrule
  BM24 & " &  $90$ & $200$ & $300$ & $10$ & $0.2$ & $801.98$ & $0.263$ & $0.258$ & $1.06$ & $0.3$ & $-1.35$ & ${ 141.06  }$ & 1.1 & $42$  \\
 \bottomrule
		\end{tabular}}

	\noindent{\footnotesize{\textsuperscript{$\ddagger$} It means the present value follows the previous value in the same column.}}
\end{table}

As mentioned above, and now evident from the Table \ref{Table:1 Benchmark values} and Figure \ref{2hdm-entropy-fig} the dilution is more intense in the scenario when multiple higgs fields are involved.

\section{Conclusions}
 As shown in this paper, the preexisting baryon asymmetry, and the frozen out {DM} density of the universe can be diluted significantly due to electroweak phase transition. In particular, such dilution can take place for DM particles with the mass $m\gg 5$~TeV, particles interacting with plasma with cross-section $\sigma \le (T_c M_{Pl})^{-1}$ and primordial black holes (PBH) with mass (see, e.g., \cite{PBHrev} for review and references) $$M< M_{Pl}\lt(\frac{M_{Pl}}{T_c}\rt)^2 \approx 2  \times 10^{28}\g \approx 10^{-5} M_{\odot}.$$ 
%
($5.6\times 10^{28}$~{$gram$} 
 or $3\times 10^{-5}M_{\odot}$ for SM and $3\times 10^{28}$~{$gram$} or $1.5\times 10^{-5}$~{$M_{\odot}$} for 2HDM ($T_c= 245~${$GeV$})).
%
These types of DM are frozen out, decoupled, frozen in, or formed before EWPT. Their dilution is achieved with the influx of entropy into the primordial plasma, which was in the state of thermal equilibrium before the onset of EWPT. We have considered two scenarios. First, in the SM scenario, we have assumed EWPT is of second order and we see the dilution factor is $\sim$13\%. {Second,} we have considered the simplest extension of SM, namely 2HDM. From Table \ref{Table:1 Benchmark values}, we see a significant difference between the two scenarios. Figures \ref{rat-F2-FLN} and \ref{2hdm-entropy-fig} also show major differences. This is because more entropy influx happens for first order phase transition which is the case for 2HDM.

{ One should note that such popular DM candidates, as WIMPs, QCD axion or Axion Like Particles (ALP) may avoid the considered effect of entropy inflow at the EWPT. Indeed, WIMPs with mass, less than $f T_c$ (where the numerical factor $f \sim 20$)  freeze out after EWPT, while QCD axion condensate is formed at QCD phase transition and the condensate of ALP with masses of order $10^{-20}$ eV even much later. The existence of these popular DM candidates is, however, still not experimentally verified and turning to such DM candidates as sterile neutrinos, which decouple before EWPT or superheavy gravitino, created just after inflation (see, e.g., \cite{khlopovPPNP} for review and references), we find that effect of entropy inflow at EWPT would be significant for detailed descriptions of the corresponding cosmological scenarios. We may conclude that the entropy inflow effect at EWPT may play important role in all the scenarios with superheavy and super-weakly interacting DM candidates.}

{DM} candidates, which can be diluted in EWPT, are predicted in much wider extensions of SM, than we have considered here. It would imply with necessity self-consistent treatment of effect of dilution with the account for all the specific model-dependent consequences of physics of early Universe, which are predicted in each corresponding scenario. However, the results of the present paper have demonstrated that even in the case of minimal extension of SM dilution in the EWPT of primordial {DM} or PBH density is significant and deserves special study for the confrontation of {DM} models with observations.

\section*{Acknowledgement}
The work of S.P.
and A.C. is funded by RSF Grant 19-42-02004. The research by M.K. was supported by the Ministry of Science and Higher Education of the Russian Federation under Project ``Fundamental problems of cosmic rays and dark matter'', No. 0723-2020-0040.


\begin{thebibliography}{99}
\bibitem{1}
Ryden, B. {Introduction to Cosmology}; Addison-Wesley: San Francisco, CA, USA, 2003; ISBN 0-8053-8912-1.  
https://doi.org/10.1017/9781316651087

\bibitem{2}
Jonathan, A. {Quarks, Leptons, and the Big Bang}; CRC Press: Boca Raton, FL, USA, 2016;  ISBN 978-0-7503-0806-9. 
https://doi.org/10.1201/9781315381367



\bibitem{3}
Caprini, C.; Chala, M.; Dorsch, G.C.; Hindmarsh, M.; Huber, S.J.; Konstandin, T.; Kozaczuk, J.; Nardini, G.; No, J.M.; \mbox{Rummukainen, K.;} et al. Detecting gravitational waves from cosmological phase transitions with LISA: An update. { J. Cosmol. Astropart. Phys.} {\bf 2020}, {3}, 024,
https://doi.org/10.1088/1475-7516/2020/03/024.

\bibitem{4}
Ghiglieri, J.; Jackson, G.; Laine, M.; Zhu, Y.
Gravitational wave background from Standard Model physics: Complete leading order. { J. High Energy Phys.} {\bf 2020}, {7}, 092. https://doi.org/10.1007/JHEP07(2020)092. 

\bibitem{5}
Sakharov, A.D. Violation of CP Invariance, C Asymmetry, and Baryon Asymmetry of the Universe. { J. Exp. Theor. Phys. Lett.}  {\bf 1967}, {5}, 24--27.
https://doi.org/10.1070/PU1991v034n05ABEH002497

\bibitem{Gorbunov:2011zzc}
Gorbunov, D.S.; Rubakov, V.A.
{Introduction to the Theory of the Early Universe: Cosmological Perturbations and Inflationary Theory}; {World Scientific}: Singapore, 2011. https://doi.org/10.1142/7874.

\bibitem{Bambi:2015mba}
Bambi, C.; Dolgov, A.D. {Introduction to Particle Cosmology}; Springer: Berlin/Heidelberg, Germany, 2015. 
https://doi.org/10.1007/978-3-662-48078-6.


\bibitem{Chaudhuri:2021agl}
Chaudhuri, A.; Khlopov, M.Y. 
Entropy production due to electroweak phase transition in the framework of two Higgs doublet model.
{ Physics}  {\bf 2021}, {3}, 275. https://doi.org/10.3390/physics3020020.


\bibitem{Dolgov:2000ht}
Dolgov, A.D.; Naselsky, P.D.; Novikov, I.D.
Gravitational waves, baryogenesis, and dark matter from primordial black holes.  { arXiv} {\bf 2000},
 arXiv:astro-ph/0009407.



\bibitem{Chaudhuri:2020wjo}
Chaudhuri, A.; Dolgov, A. PBH evaporation, baryon asymmetry,~and dark matter.
{ arXiv} {\bf 2020}, arXiv:2001.11219.



\bibitem{Schettler:2010wi}
Boeckel, T.; Schettler, S.; Schaffner-Bielich, J.
The Cosmological QCD Phase Transition Revisited.
{ Prog. Part. Nucl. Phys.} {\bf 2011}, {66}, 266--270. 
https://doi.org/10.1016/j.ppnp.2011.01.017
 
 \bibitem{khlopovPPNP} 
 Khlopov, M.Y. What comes after the Standard model?
 { Prog. Part. Nucl. Phys.} {\bf 2021}, {116} 
 103824. 
 https://doi.org/10.1016/j.ppnp.2020.103824
 
 \bibitem{Chaudhuri:2021ppr}
 Chaudhuri, A.; Khlopov, M.Y.
 Balancing Asymmetric Dark Matter with Baryon Asymmetry and Dilution of Frozen Dark Matter by Sphaleron Transition.
 { Universe} {\bf 2021}, {7}, 275.
 https://doi.org/10.3390/universe7080275.



\bibitem{Chaudhuri:2021rwt}
Chaudhuri, A.; Khlopov, M.Y.; Porey, S. 
Effects of 2HDM in Electroweak Phase Transition.
{ Galaxies} \textbf{2021}, {9}, 45. 
https://doi.org/10.3390/galaxies9020045



\bibitem{Chaudhuri:2021vdi}
Chaudhuri, A.; Khlopov, M.Y.; Porey, S.
Entopy release in Electroweak Phase Transition in 2HDM. { arXiv} \textbf{2021},
arXiv:2111.00139.



\bibitem{1306.2314}
Viel, M.; Becker, G.D.; Bolton, J.S.; Haehnelt, M.G. Warm dark matter as a solution to the small scale crisis: New constraints from high redshift Lyman-\ensuremath{\alpha} forest data. { Phys. Rev. D} {\bf 2013}, \textit{88}, 043502. https://doi.org/10.1103/PhysRevD.88.043502.


\bibitem{Carr}
Carr, B.; Kühnel, F. Primordial black holes as dark matter: Recent developments. { Ann. Rev. Nucl. Part. Sci.} {\bf 2020}, \textit{70}, 355--394.
https://doi.org/10.1146/annurev-nucl-050520-125911

\bibitem{PBHrev}
 Khlopov, M.Y. Primordial black holes. {\it Res. Astron. Astrophys.} {\bf 2010}, \textit{10}, 495--528. https://doi.org/10.1088/1674-4527/10/6/001.

\bibitem{2105.01406}
Chadha-Day, F.; Ellis, J.; Marsh, D.J. Axion Dark Matter: What is it and Why Now? { arXiv} {\bf 2021}, arXiv:2105.01406.

\bibitem{ALP}
Visinelli, L. Light axion-like dark matter must be present during inflation. { Phys. Rev. D} {\bf 2017}, \textit{96},  023013. 
https://doi.org/10.1103/PhysRevD.96.023013.

\bibitem{1903.03026}
Schumann, M. Direct detection of WIMP dark matter: Concepts and status. { J. Phys. G Nucl. Part. Phys.} {\bf 2019}, \textit{46}, 103003. https://doi.org/10.1088/1361-6471/ab2ea5.

\bibitem{astro-ph/0511796}
Khlopov, M.Y. Composite dark matter from 4th generation. { Pisma Zh. Eksp. Teor. Fiz.} {\bf 2006}, \textit{83}, 3--6. 
https://doi.org/10.1134/ S0021364006010012.

\bibitem{Ibarra}
Ibarra, A. Neutrinos and dark matter. { AIP Conf. Proc.} {\bf 2015}, \textit{1666}, 140004. 
https://doi.org/10.1063/1.4915588.

\bibitem{Sidhu da}
Pandey, S.; Karmakar S.; Rakshit S. Interactions of astrophysical neutrinos with dark matter: A model building perspective. \mbox{{ J. High Energy Phys.}} {\bf 2019}, \textit{1}, 95. https://doi.org/10.1007/JHEP11(2021)215.



\bibitem{fermionic DM}
Lee, K.Y.; Kim, Y.G.; Shin, S. Singlet fermionic dark matter. { J. High Energy Phys.} {\bf 2008}, \textit{5}, 100.  
https://doi.org/10.1088/1126-6708/2008/05/100.

\bibitem{gravitino}
Steffen, F.D. Gravitino dark matter and cosmological constraints. { J. Cosmol. Astropart. Phys.} {\bf 2006}, \textit{9},  1. 
https://doi.org/10.1088/1475-7516/2006/09/001.

\bibitem{DM during ewpt}
Falkowski, A.;  No, J.M. Non-thermal dark matter production from the electroweak phase transition: Multi-TeV WIMPs and “baby-zillas”. {  J. High Energy Phys.} {\bf 2013}, \textit{2}, 34. https://doi.org/10.1007/JHEP02(2013)034.


\bibitem{Fluid}
Peebles, P.J.E. Fluid dark matter. { Astrophys. J. Lett.} {\bf 2000}, \textit{534}, L127. https://doi.org/10.1086/312677.






\bibitem{Bochkarev:1990fx}
Bochkarev, A.I.; Kuzmin, S.V.; Shaposhnikov, M.E.
Electroweak baryogenesis and the Higgs boson mass problem.
{ Phys. Lett. B} {\bf 1990}, {244}, 275--278. https://doi.org/10.1016/0370-2693(90)90069-I.
 
 
 
 


\bibitem{Logan:2014jla}
Logan, H.E. 
TASI 2013 lectures on Higgs physics within and beyond the Standard Model. {{arXiv}} {\bf 2013}, 
arXiv:1406.1786.
 
 
 
 
 
 
\bibitem{Melo}
Melo, I. Higgs potential and fundamental physics. { Eur. J. Phys.} {\bf 2017}, {38}, 065404. https://doi.org/10.1088/1361-6404/aa8c3d.

\bibitem{EWBG}
Trodden, M. Electroweak baryogenesis. {  Rev. Mod. Phys.}  {\bf 1999}, {71}, 1463.
https://link.aps.org/doi/10.1103/RevModPhys.71.1463

\bibitem{Branco:2011iw}
Branco, G.C.; Ferreira, P.M.; Lavoura, L.;  Rebelo, M.N.; Sher, M.; Silva, Joao, P. Theory and phenomenology of two-Higgs-doublet models. { Phys. Rep.}  {\bf 2021}, {516}, 1--102.
https://doi.org/10.1016/j.physrep.2012.02.002

\bibitem{Karmakar}
Karmakar, S.; Rakshit, S. 
Effective Route Beyond the Extended Scalar Sectors of the Standard Model. Doctoral Dissertation, Discipline of Physics, IIT Indore, Indian Institute of Technology Indore, Indore, India, 2020.

\bibitem{Karmakar:2020mds}
Karmakar S.
Relaxed Constraints on Masses of New Scalars in 2HDM.
{ Springer Proc. Phys.} {\bf 2020}, {248}, 193--198. 
https://doi.org/10.1007/978-981-15-6292-1\_23.


\bibitem{Bernon:2017jgv}
Bernon, J.; Bian, L.; Jiang, Y.
A new insight into the phase transition in the early Universe with two Higgs doublets.
{ J. High Energy~Phys.} {\bf 2018}, {5},~151.
https://doi.org/10.1007/JHEP05(2018)151




\bibitem{Basler:2018cwe}
Basler, P.; M\"uhlleitner, M.
BSMPT (Beyond the Standard Model Phase Transitions): A tool for the electroweak phase transition in extended Higgs sectors.
{ Comput. Phys. Commun.} {\bf 2019}, {237}, 62--85.
https://doi.org/10.1016/j.cpc.2018.11.006


\bibitem{Basler:2020nrq}
Basler, P.; M\"uhlleitner, M.; M\"uller, J.
BSMPT v2 A Tool for the Electroweak Phase Transition and the Baryon Asymmetry of the Universe in Extended Higgs Sectors.
{ arXiv} {\bf 2020},  arXiv:2007.01725.







\end{thebibliography}
\end{document}